\documentclass[3p,times,twocolumn]{elsarticle}

\usepackage{ecrc}
\usepackage{amsmath}

\volume{00}

\firstpage{1}

\journalname{Nuclear Physics B Proceedings Supplement}

\runauth{S. Fink}

\jid{nuphbp}

\jnltitlelogo{Nuclear Physics B Proceedings Supplement}

\usepackage{amssymb}

\usepackage[figuresright]{rotating}
\usepackage{color}
\usepackage{helvet}
\newcommand{\tHq}{$\mathrm{tHq}$~}
\newcommand{\ttbar}{$\mathrm{t\bar{t}}$~}

\begin{document}

\begin{frontmatter}



\dochead{}

\title{Search for H to $ \mathrm{b\bar{b}}$ in association with single top quarks as a test of Higgs boson couplings}


\author{Simon Fink}

\address{}

\begin{abstract}
\noindent The production of a Higgs boson in association with a single top quark is one of a few channels which are sensitive for not yet excluded anomalous couplings of the Higgs boson to fermions. Multivariate analysis tools are used for the reconstruction and classification of signal events, where the Higgs boson decays into bottom quarks and the single top quark decays leptonically. In this conference report the most recent results at time of the conference using the full dataset recorded by the CMS detector at 8 TeV are presented.
\end{abstract}


\end{frontmatter}


\section{Introduction}
\label{sec:Introduction}
\noindent Since the discovery of a new particle by the CMS and ATLAS collaborations in 2012~\cite{CMS-PAPERS-HIG-12-036,ATLASHiggsObs}, all measurements are found to be consistent with the Higgs Boson predicted by the standard model (SM). In order to find any deviations from the SM and therefore possible signs of new physics beyond the SM, the properties of the particle have to be measured with ever increasing precision. 
One of these properties is the value of $y_t$, the Yukawa coupling of the Higgs boson to top quarks.  Most measurements probing this property are only sensitive to the absolute value of $y_t$, rather than its sign, as quantities such as the rate of Higgs boson production in association with top-quark pairs depend only on  $|y_t|^2$.  
The cross section for the production of Higgs bosons in association with single top quarks is particularly sensitive to the sign of $y_t$, or, more precisely, the relative phase of the couplings of the Higgs to the top quark and to W bosons. 
The two amplitudes lead to a nearly total destructive interference, with an expected cross section of 18.3\,fb at 8 TeV in the standard model~\cite{Farina}.  Any alteration of the relative phase of the two amplitudes would result in a enhancement of the cross section and this process is therefore very sensitive to enhancements in models beyond the standard model.

\section{Analysis overview}
\label{sec:analysis}
\noindent The analysis which is described here focusses on the search for a signal process under the hypothesis of $y_t = -1$. Due to constructive interference the production cross section of this process would be enhanced by a factor of $\sim13$ up to $\sigma = 234$ fb.
\subsection{Event Topology and Event Selection}
\label{subsec:selection}
\begin{figure}
\centering
\includegraphics[width=0.3\textwidth]{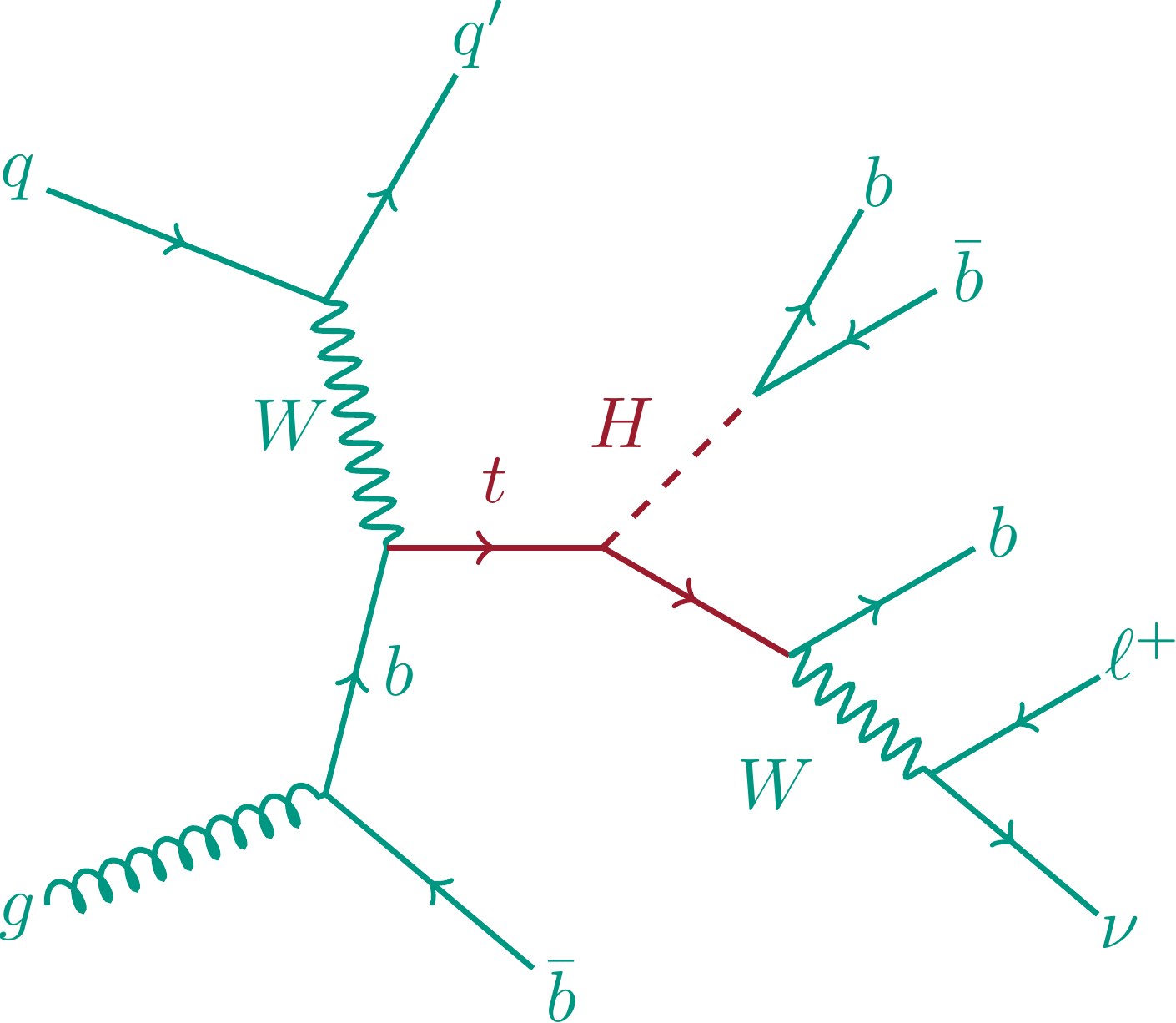}
\caption{A characteristic Feynman diagram for the final state studied in this analysis.}
\label{fig:topology}
\end{figure}

\noindent The final topology for the searched process consists of one isolated lepton, missing transverse energy, four bottom quarks as well as one distinct light forward jet. The Feynman graph for this process can be seen in Figure~\ref{fig:topology}.
\noindent In order to enhance the signal fraction in all considered events  we apply a selection where one tightly isolated lepton, either a electron or a muon, is required; the missing transverse energy due to the escaping neutrino has to be larger than 45 GeV in the electron channel and larger than 35 GeV in the muon channel and jets have to have a transverse momentum greater than 20 GeV, if they lie in the central part of the detector ($|\eta| < 2.4$), and greater than 40 GeV, if they lie in the outer detector regions ($|\eta|>2.4)$ . One of these jets has to be untagged, and at least 4 jets have to have a transverse momentum greater than 30 GeV. The remaining events are sorted according to the multiplicity of b-tagged jets into two independent signal regions and a control regions, events with three b-tagged jets form the three tag (3T) signal region, events with four b-tagged jets form the four tag (4T) region and events with two tags are sorted into the control region (2T) which is predominantly populated by \ttbar events.
\subsection{Top pair production modelling}
\noindent As mentioned before top pair production is the dominant background in this analysis  and two approaches are used to model it. One approach is relying completely on the provided Monte Carlo simulation of this process, the alternative approach is using a data-driven method trying to model the process in our signal region based on the kinematics seen in the control region.\\
Events from the two tag control region are given a weight according to the probability of  an event with the same kinematics  to appear the 3T or 4T region.  Consider an event with $n$ reconstructed jets in which each jet has a unique b-tagging efficiency, $\epsilon (p_{i},f_{i})$, that is dependent on the transverse momentum and flavor of the jet.  The probability of a selected event (but for the number of b tags) to have $m$ tagged jets is
\begin{equation}
 P_{m} = \sum_{comb} \prod_{i=1}^m \epsilon (p_{i},f_{i}) \cdot \prod_{j=m+1}^n (1-\epsilon(p_{j},f_{j})) .
\end{equation}
The final weight for each event is then calculated in the following way:
\begin{equation}
 w = P_{3}/P_{2} \quad \mathrm{or} \quad P_{4}/P_{2}
\end{equation}
\subsection{Event Reconstruction} 
\noindent The small signal-to-background ratio after the selection makes it essential to exploit multivariate techniques to discriminate signal processes from background processes. In order to construct observables which have a good separation power, we employ two dedicated "jet-assignment MVAs", where the jets in an event are assigned to the quarks in the final state of a matrix element. One of those MVAs matched the jets under the hypothesis that the event is a signal \tHq event and the second one matches the jets under the hypothesis that the event is a \ttbar event. Reconstructing all events under these two hypothesis allows the construction of variables separating signal especially well against the dominant \ttbar background.\\
The higher the multiplicity of jets in an event is, the more possibilities arise to assign jets to the quarks of the specific process. Therefore we apply constraints reducing the number of jet assignment possibilities. 
\subsection{Event Classification}
\noindent Events are then classified as either signal or background by employing a third MVA, which uses three categories of input variables. The first category consists of observables which are not dependent on any of the jet assignments done before and exploit global properties of an event. The second category exploits observables obtained with the \tHq reconstruction and the third category uses observables obtained with the \ttbar reconstruction. A total number of eight input variables are used to separate signal from background. In Figure 2 the eight different input variables can be seen sorted into the three different categories. As an example the most discriminating variable, the absolute value of the pseudorapidity of the reconstructed light jet, is shown in Figure 3.
The response distributions of the classification MVA for signal and background events can be seen in Figure 4. A good separation can be achieved.
 \begin{figure}[h!]
 \centering
 \label{tab:results}
 \includegraphics[width=0.45\textwidth]{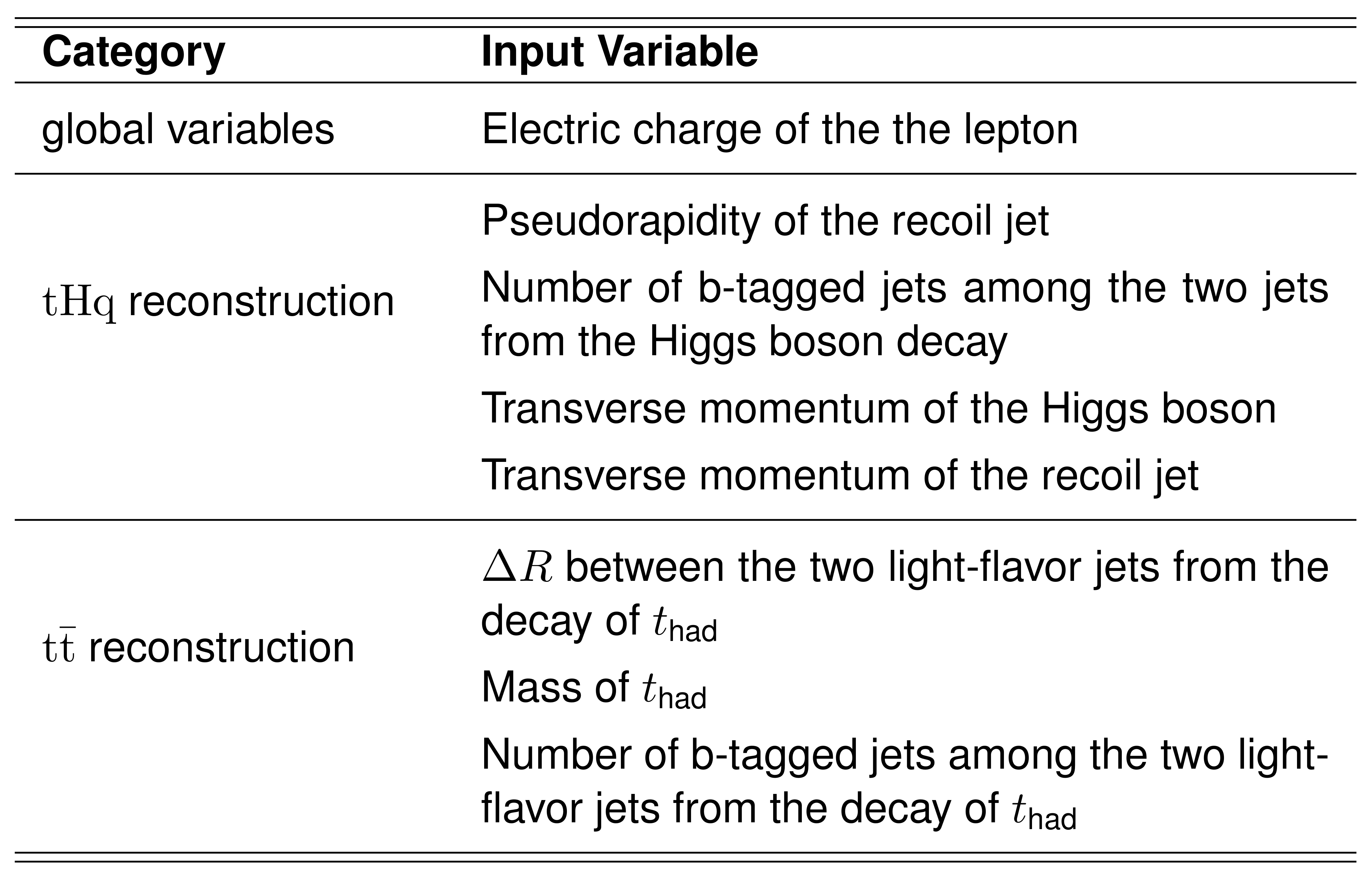}
 \caption{Input variables for the classification MVA. The variables are split into three groups: global variables, variables of the jet assignment under the \tHq hypotheses, variables of the jet assignment under the \ttbar hypothesis. In the descriptions, $\mathrm{t}_{had}$ stands for a hadronically decaying top quark.}
 \end{figure}
 \begin{figure}
 \centering
\label{fig:postfit}
 \includegraphics[width=0.45\textwidth]{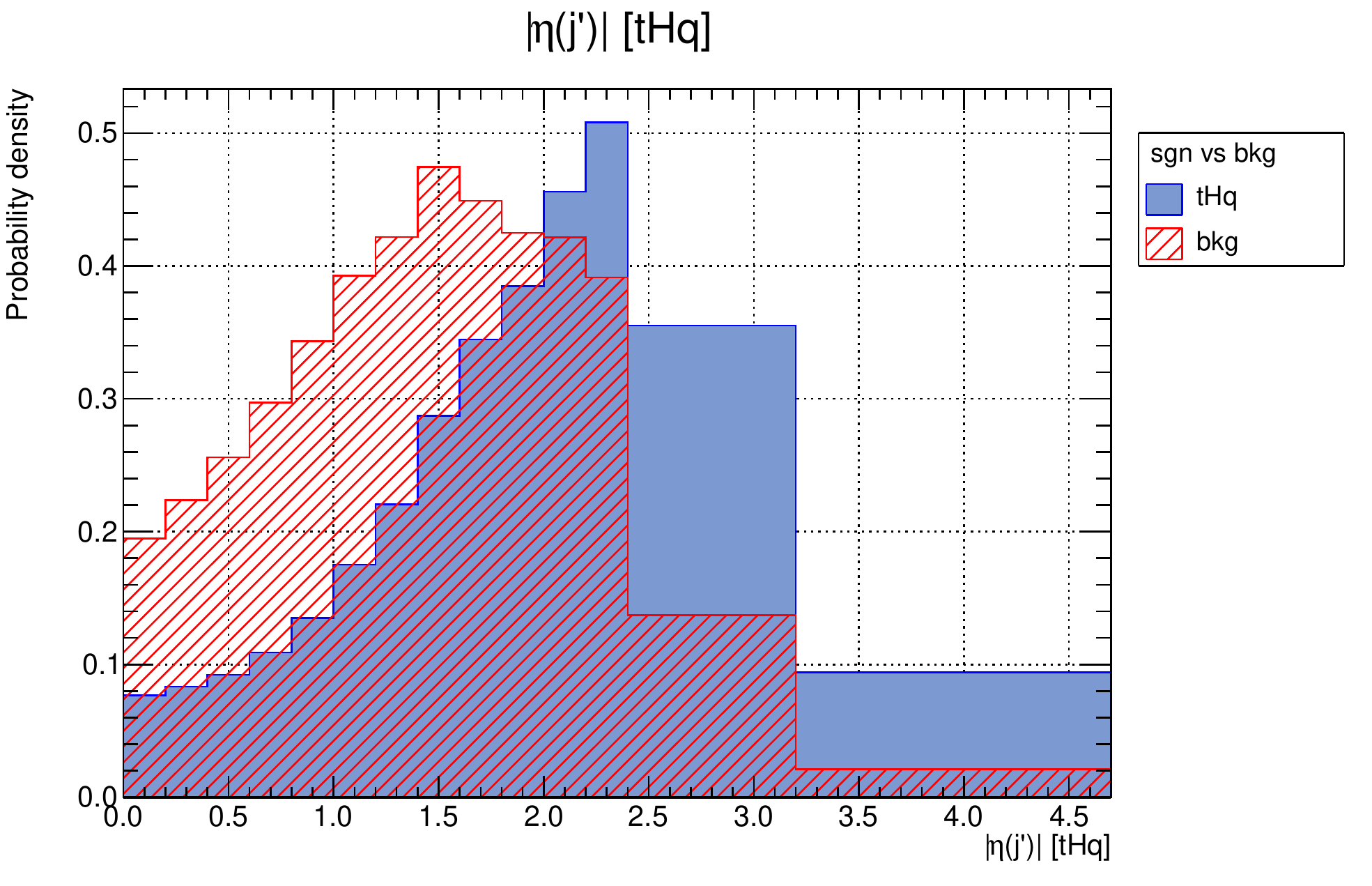}
  \put(-85,110){\textcolor{red}{ \tiny  \fontfamily{phv} \selectfont  Work in progress}}
 \caption{Distributions of signal and background events over the absolute value of the pseudorapidity of the reconstructed light jet, which is used as input variable for the classification MVA. This variable is the single most discriminating variable between signal and background events.}
 \end{figure}

\begin{figure}[b!]
\centering
\label{fig:MVAout}
 \includegraphics[width=0.45\textwidth]{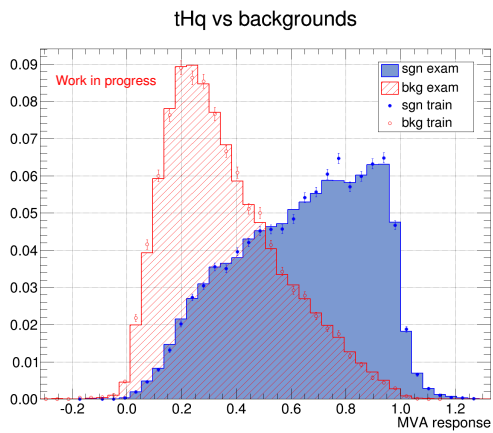}
 \put(-185,135){\textcolor{red}{ \tiny  \fontfamily{phv} \selectfont  Work in progress}}
 \caption{Distributions of the classification MVA’s response for signal and background events. A good separation has been achieved.}
 \end{figure}
 \section{Summary}
\noindent With the full dataset at a center of mass energy of 8 TeV a search for t-channel single top production in association with the 125 GeV Higgs boson is conducted. The \tHq decay channel is used with H $\to \mathrm{b\bar{b}}$. The signal is assumed to come from the $y_{t} = -1 $ coupling which would require beyond the standard model physics.
The assignment of the jets is using two separate artificial neural networks. This analysis pioneers the technique, that each event is reconstructed under a potential \tHq signal hypothesis and a \ttbar background hypothesis. The individual reconstructions provide input variables with a high separation power between signal and background events.  A third neural network then uses these input variables, obtained from the reconstructions, to enhance the signal-to-background ratio and discriminate against the main backgrounds, including the predominant \ttbar production. This process is modelled with a simulation-driven approach and a data-driven approach cross-checking each other. The analysis was not yet public at the time of the conference, but results have been made public shortly afterwards \citep{HIG14015}.




\nocite{*}
\bibliographystyle{elsarticle-num}
\bibliography{ichep_sfink}

\begin{thebibliography}{1}
\expandafter\ifx\csname url\endcsname\relax
  \def\url#1{\texttt{#1}}\fi
\expandafter\ifx\csname urlprefix\endcsname\relax\def\urlprefix{URL }\fi
\expandafter\ifx\csname href\endcsname\relax
  \def\href#1#2{#2} \def\path#1{#1}\fi

\bibitem{CMS-PAPERS-HIG-12-036}
S.~Chatrchyan, et~al., Observation of a new boson at a mass of {125 GeV} with
  the {CMS} experiment at the {LHC}, J. High Energy Phys. 06 (2013) 081.
\newblock \href {http://dx.doi.org/10.1007/JHEP06(2013)081}
  {\path{doi:10.1007/JHEP06(2013)081}}.

\bibitem{ATLASHiggsObs}
G.~Aad, et~al., {Observation of a new particle in the search for the Standard
  Model Higgs boson with the ATLAS detector at the LHC}, Phys.Lett. B716 (2012)
  1--29.
\newblock \href {http://arxiv.org/abs/1207.7214} {\path{arXiv:1207.7214}},
  \href {http://dx.doi.org/10.1016/j.physletb.2012.08.020}
  {\path{doi:10.1016/j.physletb.2012.08.020}}.

\bibitem{Farina}
M.~Farina, F.~Grojean, F.~Maltoni, E.~Salvioni, A.~Thamm, Lifting degeneracies
  in higgs couplings using single top production in association with a higgs
  boson, JHEP 1305 (2013) 022.
\newblock \href {http://arxiv.org/abs/1211.3736} {\path{arXiv:1211.3736}},
  \href {http://dx.doi.org/10.1007/JHEP05(2013)022}
  {\path{doi:10.1007/JHEP05(2013)022}}.

\bibitem{HIG14015}
S.~Chatrchyan, et~al., {Search for H to bbbar in association with single top
  quarks as a test of Higgs couplings}~(CMS-PAS-HIG-14-015).

\end{thebibliography}







\end{document}